\documentclass[aip,jcp,reprint,superscriptaddress,floatfix]{revtex4-1}

\usepackage[utf8]{inputenc}
\usepackage[toc,page]{appendix}
\usepackage{amsmath, amssymb}
\usepackage{braket}
\usepackage{hyperref}
\usepackage[final]{changes}
\usepackage{pgfplots}
\usepackage{verbatim}
\usepackage{bbold}
\usepackage{commath}
\usepackage{epstopdf}
\usepackage{chemformula}
\usepackage[subpreambles=true]{standalone}

\pgfplotsset{compat=newest}

\usetikzlibrary{external}

\newcommand{\rrangle}{\rangle\!\rangle}
\newcommand{\llangle}{\langle\!\langle}

\newcommand \mycomment[1]   {{\color{red} [{\it {#1}}]}}
\renewcommand{\Re}{\text{Re}}
\renewcommand{\Im}{\text{Im}}

\begin{document}

\title{Numerical stability of time-dependent coupled-cluster methods for many-electron dynamics
in intense laser pulses}
\author{H{\aa}kon Emil Kristiansen}
\email{h.e.kristiansen@kjemi.uio.no}
\affiliation{Hylleraas Centre for Quantum Molecular Sciences,
Department of Chemistry, University of Oslo,
P.O. Box 1033 Blindern, N-0315 Oslo, Norway}
\author{{\O}yvind Sigmundson Sch{\o}yen}
\email{o.s.schoyen@fys.uio.no}
\affiliation{Department of Physics, University of Oslo,
N-0316 Oslo, Norway}
\author{Simen Kvaal}
\email{simen.kvaal@kjemi.uio.no}
\affiliation{Hylleraas Centre for Quantum Molecular Sciences,
Department of Chemistry, University of Oslo,
P.O. Box 1033 Blindern, N-0315 Oslo, Norway}
\author{Thomas Bondo Pedersen}
\email{t.b.pedersen@kjemi.uio.no}
\affiliation{Hylleraas Centre for Quantum Molecular Sciences,
Department of Chemistry, University of Oslo,
P.O. Box 1033 Blindern, N-0315 Oslo, Norway}
\date{\today}

\begin{abstract}
We investigate the numerical stability of time-dependent coupled-cluster theory for many-electron dynamics in intense laser pulses, comparing two coupled-cluster formulations with full configuration interaction theory.
Our numerical experiments show that orbital-adaptive time-dependent coupled-cluster doubles (OATDCCD) theory offers significantly improved
stability compared with the conventional Hartree-Fock-based time-dependent coupled-cluster singles-and-doubles (TDCCSD) formulation. The improved stability stems from greatly reduced oscillations in the doubles amplitudes,
which, in turn, can be traced to the dynamic biorthonormal reference determinants of OATDCCD theory. As long as these are good approximations to the Brueckner determinant, OATDCCD theory is numerically stable. We propose the reference weight as a diagnostic quantity to identify situations where the TDCCSD and OATDCCD theories become unstable.
\end{abstract}

\maketitle

\section*{Introduction}
With the advent of ultrashort, intense laser pulses capable of probing electronic processes with high resolution in
both space and time,~\cite{Peng2019}
the demand for highly accurate simulations of many-electron dynamics is increasing.
The most widely used wave function-based method is multiconfiguration time-dependent Hartree-Fock (MCTDHF) theory (see, e.g.,
Ref.~\onlinecite{Meyer2009}) which, unfortunately, quickly becomes prohibitively expensive as the number of electrons grows.
The more benignly scaling coupled-cluster (CC) hierarchy of methods~\cite{Bartlett2007} offers an alternative, which
has only recently been explored in the context of laser-driven many-electron
dynamics.~\cite{Huber2011,OATDCC,HeadGordon_2016,Nascimento2016,Nascimento2017,TDOCC_2018,Symplectic_TDCC_2018,RelRTCC_2019, Nascimento2019}

\citeauthor{Symplectic_TDCC_2018}~\cite{Symplectic_TDCC_2018} showed that
time-dependent CC (TDCC) theory,
formulated with the static Hartree-Fock (HF) 
reference determinant, becomes numerically challenging in the presence of strong laser pulses.
The numerical issues arise as the HF determinant becomes a poor reference function for the 
TDCC state vector. This 
occurs when
a 
laser pulse pumps the many-electron system into a
state with very low HF weight, causing large and sudden changes in the
amplitudes, which require infeasibly tiny time steps in the numerical integration. The instability was observed even for a two-electron system where TDCC singles-and-doubles (TDCCSD) is formally exact.

The instability
resembles
the multireference problem in CC theory, which tends to be
accompanied by unusually large doubles amplitudes.~\cite{Giner2018}
The action of a laser pulse, however, is represented semiclassically by a one-electron operator, whose main effects should be capturable by a \emph{moving} reference determinant.
We have explored the original TDCC method,~\cite{Hoodbhoy_Negele} which is based on the time-dependent HF reference determinant, and found that it
does not cure the instabilities.
Brueckner CC theory~\cite{Handy1989} is not an attractive solution, as it leads to spurious pole structures in nonlinear response functions.~\cite{Aiga1994}
Time-dependent orbital-optimized CC (TDOCC)~\cite{NOCC_1999} theory has been used 
to simulate high-harmonic generation and ionization of the argon atom.~\cite{TDOCC_2018}
While TDOCC theory thus appears to be stable, 
it does not converge to the full configuration-interaction
(FCI) limit for 
more than two electrons.~\cite{OCC_NOT_FCI}
The correct FCI limit is achieved,~\cite{NOCC_FCI} however, with
time-dependent nonorthogonal orbital-optimized
CC (TDNOCC)~\cite{NOCC_2001} theory
and
orbital-adaptive time-dependent
CC (OATDCC)~\cite{OATDCC} theory.

In this work we investigate the stability of 
OATDCC doubles (OATDCCD) theory,
restricting ourselves to simulations that allow for comparison with time-dependent FCI (TDFCI) theory. 

\section*{Theory}
Following 
Ref.~\onlinecite{Symplectic_TDCC_2018},
the TDCC and OATDCC \emph{Ans{\"a}tze}
for the quantum state of a many-electron system 
can be written
as
\begin{equation}
\vert S(t) \rrangle = \frac{1}{\sqrt{2}}
   \begin{pmatrix}
      \vert \Psi(t)\rangle \\
      \vert \tilde{\Psi}(t)\rangle
   \end{pmatrix},
\label{eq:S}
\end{equation}
where $t$ denotes time, and
\begin{align}
   &\vert \Psi(t)\rangle
   =
   \text{e}^{T(t)}\vert \Phi_0(t)\rangle \text{e}^{\tau_0(t)},
   \\
   &\langle \tilde{\Psi}(t) \vert
   = \text{e}^{-\tau_0(t)} \langle \tilde{\Phi}_0(t)\vert (\lambda_0(t) + \Lambda(t))\text{e}^{-T(t)},
\end{align}
such that, with $\lambda_0 = 1$, $\vert S(t) \rrangle$ is normalized with respect to the indefinite inner product~\cite{Symplectic_TDCC_2018}
\begin{equation}
    \llangle S_1 \vert S_2 \rrangle = \frac{1}{2}\langle \tilde{\Psi}_1 \vert \Psi_2\rangle
                                    + \frac{1}{2}\langle \tilde{\Psi}_2 \vert \Psi_1\rangle^*.
\end{equation}
The expectation value of an operator $P$ then becomes
\begin{equation}
     \llangle S(t) \vert \Hat{P} \vert S(t)\rrangle =
          \frac{1}{2}\langle \tilde{\Psi}(t) \vert P \vert \Psi(t)\rangle
        + \frac{1}{2}\langle \tilde{\Psi}(t) \vert P^\dagger \vert \Psi(t)\rangle^*,
\label{eq:state_vector}
\end{equation}
where $\Hat{P} = P\mathbb{1}$ with $\mathbb{1}$ the $2\times 2$ unit matrix.

In 
TDCC theory,~\cite{Symplectic_TDCC_2018},
$\vert \Phi_0(t)\rangle = \vert \Phi_\text{HF}\rangle$ is the static 
HF determinant and $\langle \tilde{\Phi}_0(t)\vert = \langle \Phi_\text{HF}\vert$. The cluster operator
$T(t)$ ($\Lambda(t)$) contains from single to $n$-tuple
excitation (de-excitation) operators with respect to the HF determinant with
$1 \leq n \leq N$, $N$ being the number of electrons. The cluster operators
are parameterized by the amplitudes $\tau(t)$ and $\lambda(t)$, one amplitude per excitation and de-excitation. While $\lambda_0(t)$ is a normalization variable, $\tau_0(t)$ is a phase variable. Both are treated as dynamical parameters on an equal footing with the correlating amplitudes.

In OATDCC theory,~\cite{OATDCC}
the whole set of determinants used to build the bra and ket components
$\langle\tilde{\Psi}(t)|$ and $|\Psi(t)\rangle$ are dynamical parameters.
Single excitations (de-excitations)
are removed from $T(t)$ ($\Lambda(T)$) as these are redundant,~\cite{NOCC_2001,OATDCC}
and the underlying spin orbitals form a biorthonormal
set, $\langle \tilde{\varphi}_p(t) \vert \varphi_q(t)\rangle = \delta_{pq}$.
We use indices $p,q,r,\ldots$ to denote general spin orbitals,
while $i,j,k,\ldots$ and $a,b,c,\ldots$ denote occupied and virtual spin orbitals, respectively, with respect to the reference determinant at time $t$.

Without truncation in the cluster operators ($n=N$),
$\vert \Psi(t)\rangle$ and $\langle \tilde{\Psi}(t) \vert$ are proportional to the TDFCI wave function and its conjugate, respectively.

Using the time-dependent bivariational principle,~\cite{Arponen1983}
the amplitude equations, with time-dependence suppressed for notational convenience, are~\cite{OATDCC}
\begin{align}
    \text{i} &\dot{\tau}_\mu = \braket{\tilde{\Phi}_\mu|e^{-T}(H-\text{i}D_0)\text{e}^{T}|\Phi_0},
\label{OATDCC_tau_equations} \\
   -\text{i} &\dot{\lambda}_\mu
   = \braket{\tilde{\Phi}_0|(1+\Lambda)e^{-T}[H-\text{i}D_0,X_\mu]\text{e}^{T}|\Phi_0},
\label{OATDCC_lambda_equations}
\end{align}
where $\mu\geq 0$, $X_\mu$ is an excitation operator such that $\ket{\Phi_\mu} = \hat{X}_\mu \ket{\Phi_0}$, $\hat{X}_0 = 1$,
$\braket{\tilde{\Phi}_\mu| \Phi_\nu} = \delta_{\mu\nu}$, and
(using Einstein's summation convention throughout)
\begin{equation}
    D_0 = \braket{\tilde{\varphi}_p|\dot{\varphi}_q}c_p^\dagger \tilde{c}_q,
\end{equation}
arises from the time-dependent orbitals.
Apart from this correction, Eqs.~\eqref{OATDCC_tau_equations} and~\eqref{OATDCC_lambda_equations} are the usual time-dependent
CC amplitude equations.
With the static HF reference determinant, $D_0 = 0$. Note that $\dot{\lambda}_0 = 0$, implying that the norms of the TDCC and OATDCC state vectors are conserved. 
The creation and annihilation operators 
$c_p^\dagger$ and $\tilde{c}_p$ result from a similarity transformation of an orthonormal set of creation and annihilation operators, and refer to the biorthonormal orbitals in OATDCC theory and to the orthonormal HF orbitals
in TDCC theory. They satisfy the usual anticommutation relations for fermions.~\cite{OATDCC,NOCC_2001}

In analogy with MCTDHF theory, OATDCC theory supports splitting of the orbital space into active and inactive
subspaces.~\cite{OATDCC}
In this work, however, all orbitals are chosen active such that~\cite{OATDCC}
\begin{equation}
    \ket{\dot{\varphi}_q} =  \ket{\varphi_p} \eta^p_q,  \qquad
    \bra{\dot{\tilde{\varphi}}_q} =  -\eta^p_q \bra{\tilde{\varphi}_q},
\end{equation}
where the nonzero components of $\boldsymbol{\eta}$ are determined from the linear equations
\begin{equation}
\text{i} A^{ib}_{aj} \eta^j_b = R^i_a, \qquad
-\text{i} A^{ja}_{bi} \eta^b_j = R^a_i.
\label{Eta eq}
\end{equation}
The right-hand sides are given 
by Eqs. (30a) and (30b) of Ref.~\onlinecite{OATDCC}, and
$A^{ib}_{aj} = \braket{\tilde{\Psi}|[c_j^\dagger \tilde{c}_b,c_a^\dagger \tilde{c}_i]|\Psi}$.


Truncating the cluster operators after doubles yields the OATDCCD and TDCCSD methods.
This simplifies the OATDCCD equations, as the operator $D_0$ drops from Eqs.~\eqref{OATDCC_tau_equations}
and~\eqref{OATDCC_lambda_equations}.
Keeping all orbitals active ($Q=0$ in the notation of Ref.~\onlinecite{OATDCC}),
the OATDCCD method becomes equivalent to TDNOCCD, allowing us to optimize the ground state as outlined in Ref.~\onlinecite{Quadratically_conv_OCC_algo}.

\citeauthor{Symplectic_TDCC_2018}~\cite{Symplectic_TDCC_2018}
argue that numerical instabilities arise when the HF determinant becomes a poor reference for the TDCCSD state vector. We thus need to quantify the quality of the reference determinant(s) in TDCCSD and OATDCCD theory. In analogy with Eq.~\eqref{eq:state_vector}, we define the reference state vector
\begin{equation}
    \vert R(t) \rrangle = \frac{1}{\sqrt{2}}\begin{pmatrix}
      \vert \Phi_0(t) \rangle \\
      \vert \tilde{\Phi}_0(t)\rangle
   \end{pmatrix},
\label{eq:Referencestate}
\end{equation}
and introduce the CC reference weight as 
\begin{equation}
    W_\text{CC} = \vert\llangle R(t) | S(t) \rrangle \vert^2 
    =
    \frac{1}{4} \vert A(t) + \tilde{A}^*(t)\vert^2.
 \label{CC reference weight}
\end{equation}
The quantities 
\begin{equation}
    A(t) = e^{\tau_0(t)}, \qquad
    \tilde{A}(t) = e^{-\tau_0(t)}(1- \tilde{B}(t)).
\end{equation}
measure the
reference weights of $\ket{\Psi(t)}$ and $\bra{\tilde{\Psi}(t)}$, respectively. If either of these is close to zero, numerical issues must be expected as observed for TDCCSD theory by \citeauthor{Symplectic_TDCC_2018}.~\cite{Symplectic_TDCC_2018}

In the TDCCSD approximation,
\begin{align}
    \tilde{B}(t) &= \frac{1}{4}\lambda^{ij}_{ab}(t)
    \left(
        \tau^{ab}_{ij}(t)
        - \frac{1}{2} P(ab) P(ij) \tau^{a}_{i}(t) \tau^{b}_{j}(t)
    \right)
    \nonumber \\
    &\qquad
    - \lambda^i_a(t)\tau^a_i(t)
\end{align}
from which the OATDCCD expression is obtained by removing the terms containing singles amplitudes.
With $\boldsymbol{M}$ an arbitrary tensor, the permutation operators are defined by
$P(pq) M^{p q\dots}_{r s \dots}
    = M^{p q \dots}_{r s \dots}
    - M^{q p \dots}_{r s \dots}$.

In order to judge the quality of a given reference determinant for single-reference CC theory, we compute its weight in the TDFCI state,
\begin{equation}
    W_\text{FCI}(t)
    = \vert\langle\Phi_0(t)\lvert\Psi_{\text{FCI}}(t)\rangle\vert^2,
\end{equation}
where we assume that $\ket{\Phi_0(t)}$ is normalized.
While the weight of the HF determinant, $\ket{\Phi_0(t)} = \ket{\Phi_\text{HF}}$, is trivially computed when the TDFCI wave function is expressed in the orthonormal HF determinant basis, it is not obvious which of the two biorthonormal reference determinants in OATDCC theory should be used.
Due to the similarity with Brueckner CC doubles theory,~\cite{Handy1989} we conjecture that the OATDCCD reference determinants approximate the
Brueckner determinant (and its conjugate). We use the term Brueckner determinant exclusively for the single Slater determinant with maximum overlap with the TDFCI state at time $t$,~\cite{Brueckner1956,Nesbet1958,Lowdin1962}
\begin{equation}
    \ket{\Phi_B(t)} \equiv \underset{\ket{\Phi}}{\arg\max}
    \braket{\Phi|\Psi_\text{FCI}(t)}.
\end{equation}
Hence, $W_\text{CC}$ should approximate the TDFCI weight of the HF determinant for TDCCSD and that of the Brueckner determinant for OATDCCD.

The FCI weight is bounded according to 
$0 \leq W_\text{FCI} \leq 1$, and the same bounds should apply to
$W_\text{CC}$.
However, as the CC phase parameter $\tau_0(t)$ is complex, we have no guarantee that $W_\text{CC}$ is bounded from above.
With $x(t)=\Re(\tau_0(t))$ and $y(t)=\Im(\tau_0(t))$, we have $|A(t)|^2 = \text{e}^{2x(t)}$ and $|\tilde{A}(t)|^2 = \text{e}^{-2x(t)}|1-\tilde{B}(t)|^2$.
If $|x(t)|$ becomes large, one of $|A(t)|^2$ or $|\tilde{A}(t)|^2$ approaches zero while the other increases exponentially, typically making $W_\text{CC}$ greater than $1$ (unless $\tilde{B}(t) \simeq 1$).
This indicates that the CC state is a poor approximation to the TDFCI wave function and numerical issues must be expected.

\section*{Numerical experiments}
Our implementations of the TDCCSD and OATDCCD theories require a backend to generate the Hamiltonian integrals and an initial set of orthonormal orbitals. In this work, we use the PySCF software framework~\cite{PySCF} and Gaussian basis sets. Our TDFCI implementation exploits the contraction algorithms available in the PySCF interface.
The cost of TDFCI theory limits the size of the systems we can consider, both in particle number and basis set size. We present results for the \ch{He} and \ch{Be} atoms, and for the \ch{LiH} molecule placed on the $z$-axis with the \ch{Li} atom at the origin and the \ch{H} atom at $z = 3.08\,a_0$.
We use the cc-pVDZ, aug-cc-pVDZ, and cc-pVTZ basis sets.~\cite{Dunning1989,Kendall1992,Woon1994,Prascher2011}

We assume the electronic system is in the ground state at $t=0\,\text{a.u.}$ and expose it to a laser pulse polarized along the $z$-axis. In the semiclassical electric-dipole approximation, the interaction operator is
\begin{equation}
    V(t) = d_zE_{\text{max}}\sin\left(\omega t + \phi\right)
    G(t),
\end{equation}
where $d_z = -z$ is the dipole moment along the $z$-axis, $E_{\text{max}}$ is the maximum field strength, $\omega$ the carrier frequency, $\phi$ the phase, and $G(t)$ is an envelope function. We use the sinusoidal envelope given by
\begin{equation}
    G(t) = \sin^2\left( \pi \frac{t}{t_d} \right) \theta(t) \theta(t_d-t)
\end{equation}
where $t_d$ is the duration of the pulse and $\theta(t)$ is the Heaviside step function. The equations of motion are propagated in time using the Gauss-Legendre (G-L) integrator~\cite{hairer2006geometric} as described in Ref.~\onlinecite{ Symplectic_TDCC_2018}. The G-L integrator is an $s$-stage implicit Runge-Kutta integration scheme of order $2s$. We use $s=3$, time step $\Delta t = 0.01\,\text{a.u.}$, and a convergence threshold $\epsilon = 10^{-5}$ for the fixed-point iterations.

\begin{figure}
    \centering
    \includestandalone[mode=buildnew]{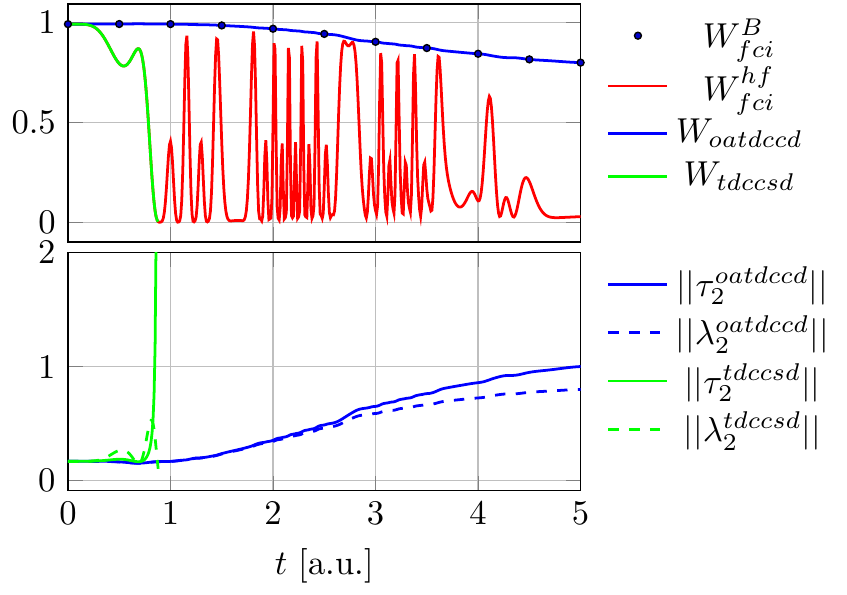}
    \caption{TDCCSD, OATDCCD and TDFCI simulations of \ch{He} with the cc-pVDZ basis exposed to a laser pulse with $E_{\text{max}}=100\,\text{a.u.}$, $\omega = 2.8735643\,\text{a.u}$, and $\phi = \pi/2$. 
    }
    \label{reference weights helium}
\end{figure}
\begin{figure}
    \centering
    \includestandalone[mode=buildnew]{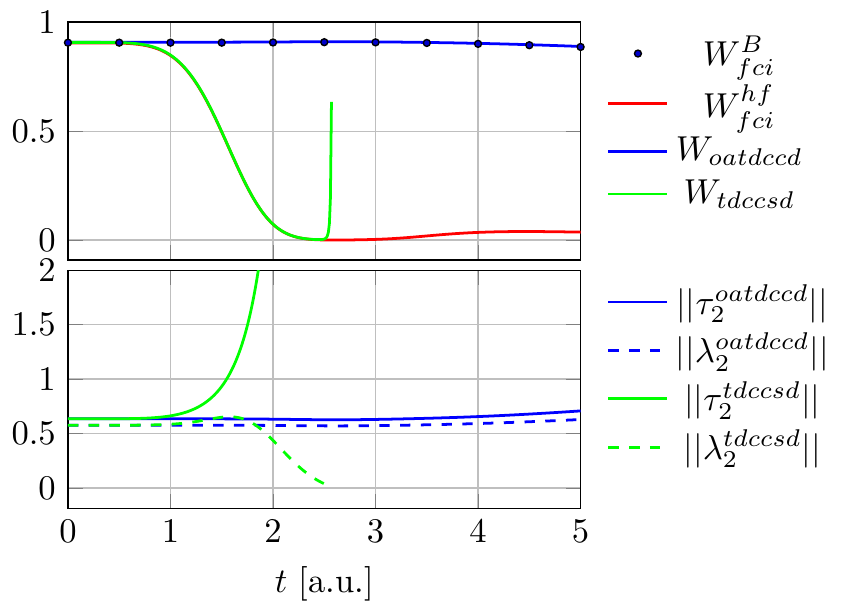}
    \caption{TDCCSD, OATDCCD and TDFCI simulations of \ch{Be} with the cc-pVDZ basis exposed to a laser pulse with $E_{\text{max}}= 1\,\text{a.u.}$, $\omega = 0.2068175\,\text{a.u}$, and $\phi = \pi/2$. 
    }
    \label{reference weights beryllium}
\end{figure}

As a first test we consider the \ch{He} and \ch{Be} simulations that caused numerical problems with the TDCCSD method in Ref.~\onlinecite{Symplectic_TDCC_2018}. The basis set is cc-pVDZ,
$t_d = 5\,\text{a.u.}$, and $\phi = \pi/2$. For \ch{He},
$E_\text{max} = 100\,\text{a.u.}$ and $\omega=2.8735643\,\text{a.u.}$. For \ch{Be}, $E_\text{max} = 1\,\text{a.u.}$ and $\omega=0.2068175\,\text{a.u.}$. The reference weights and the norm of the doubles amplitudes are
plotted in Figs.~\ref{reference weights helium} and \ref{reference weights beryllium} for \ch{He} and \ch{Be}, respectively.
As conjectured above, $W_\text{CC}$ approximates the weight of the HF and of the Brueckner determinant in the TDFCI expansion for the TDCCSD and OATDCCD methods, respectively. For the TDCCSD method, the norm of the $\tau_2$ amplitudes increases rapidly as the reference weight approaches zero, causing the simulation to fail. 

\replaced{This can in principle be handled by reducing the time step (see supplementary material). For the \ch{He} simulation,
$\Delta t = 10^{-3}\,\text{a.u.}$ is sufficient to complete the calculation. However, the \ch{Be} simulation using $\Delta t = 10^{-6}\,\text{a.u.}$ still shows spurious behavior in the induced dipole moment.}
{While this can in principle be handled by reducing the time step with concomitant increase in computational effort, 
the OATDCCD method clearly offers a more attractive alternative where the amplitude norms remain well-behaved 
throughout the simulation.}

Comparing the OATDCCD and TDFCI induced dipole moments over the entire simulation, the maximum absolute deviations were found to be on the order of $10^{-5}$ and $10^{-4}\,\text{a.u.}$ for \ch{He} and \ch{Be}, respectively. \added{Tightening the convergence parameters of the G-L integrator (see supplementary material) the discrepancy between OATDCCD and TDFCI for the \ch{He} simulation is reduced to be on the order of $10^{-10}\,\text{a.u.}$.}

Next, we consider \ch{Be} and \ch{LiH} exposed to a laser pulse for three optical cycles with $E_{\text{max}} = 0.1\,\text{a.u.}$ and phase $\phi = 0$. The carrier frequency equals the FCI excitation energy of the first dipole-allowed transition. 
For \ch{Be} we use the aug-cc-pVDZ ($\omega=0.1989\,\text{a.u.}$) and cc-pVTZ ($\omega=0.1990\,\text{a.u.}$) basis sets, while for \ch{LiH} we use the aug-cc-pVDZ ($\omega=0.1287\,\text{a.u.}$) basis set.

\begin{figure}
    \centering
    \includestandalone[mode=buildnew]{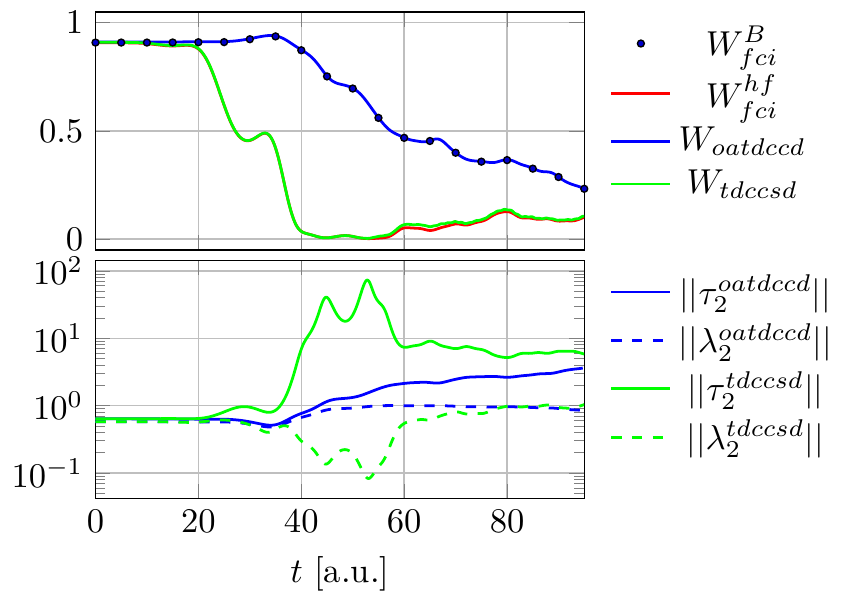}
    \caption{TDCCSD, OATDCCD and TDFCI simulations of \ch{Be} with the aug-cc-pVDZ basis exposed to a laser pulse with $E_\text{max} = 0.1\,\text{a.u.}$, $\omega = 0.1989\,\text{a.u}$, and $\phi = 0$. 
    }
    \label{reference weights beryllium aug-cc-pvdz}
\end{figure}

\begin{figure}
    \centering
    \includestandalone[mode=buildnew]{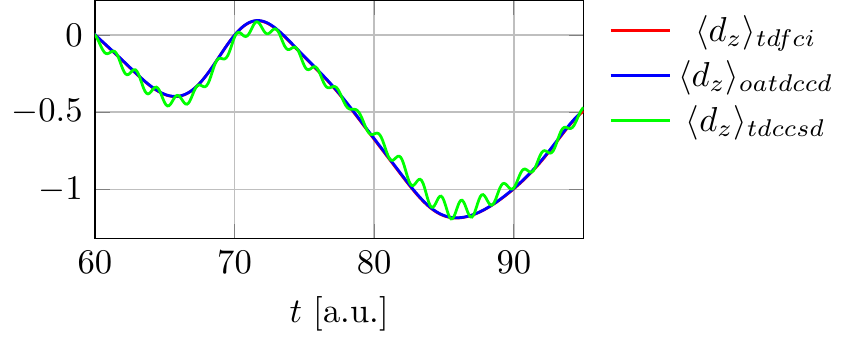}
    \caption{Time-dependent dipole moment from the simulations in 
    \autoref{reference weights beryllium aug-cc-pvdz}.}
    \label{reference weights beryllium aug-cc-pvdz-zoom}
\end{figure}

Figure \ref{reference weights beryllium aug-cc-pvdz} shows the reference weight and amplitude norms for the \ch{Be} atom with the aug-cc-pVDZ basis set. We observe that the TDCCSD and OATDCCD reference weights accurately approximate the HF and Brueckner weights in the TDFCI wave function, respectively, and that small reference weights are accompanied by large amplitude norms. While the TDCCSD simulation does not fail completely, the oscillations of the induced dipole moment, highlighted in Fig.~\ref{reference weights beryllium aug-cc-pvdz-zoom}, are not present in the TDFCI simulation, indicating numerical difficulties. The maximum absolute deviation between the TDFCI and OATDCCD dipole moment over the entire simulation is $0.005\,\text{a.u.}$. Increasing the basis set to cc-pVTZ \added{(see supplementary material)}, the TDCCSD method fails while the OATDCCD method compares well with TDFCI theory. The maximum absolute deviation between the TDFCI and OATDCCD dipole moment over the entire simulation is $0.04\,\text{a.u.}$.

\begin{figure}
    \centering
    \includestandalone[mode=buildnew]{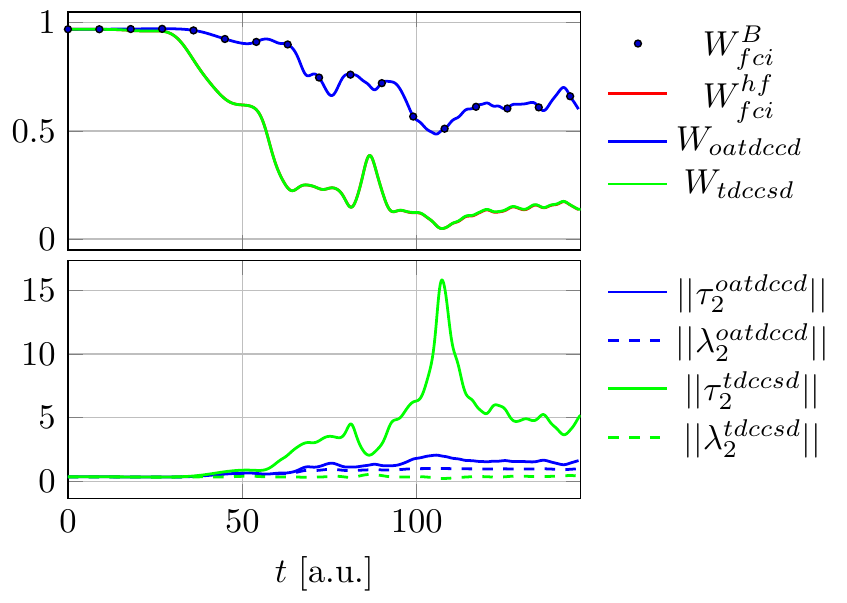}
    \caption{TDCCSD, OATDCCD and TDFCI simulations of \ch{LiH} with the aug-cc-pVDZ basis exposed to a laser pulse with $E_{\text{max}}=0.1\,\text{a.u.}$, $\omega = 0.1287\,\text{a.u}$, and $\phi = 0$. 
    }
    \label{reference weights lih aug-cc-pvdz}
\end{figure}

In the \ch{LiH} case, the TDCCSD method does not fail and the dipole moments computed with both CC methods agree well with the TDFCI result. For OATDCCD, the maximum absolute deviation in the dipole moment over the entire simulation relative to the TDFCI result is $0.015\,\text{a.u.}$, while that for TDCCSD is $0.029\,\text{a.u.}$.
We do, however, observe a sharp peak in the TDCCSD $\tau_2$ amplitude norm, shown in Fig.~\ref{reference weights lih aug-cc-pvdz}, which is absent in the OATDCCD simulation. \replaced{Increasing the field strength by a factor of two (see supplementary material), we find that the TDCCSD method breaks down.}{We thus suspect that these conditions are very close to what TDCCSD is capable of describing in a numerically stable manner.}



\begin{figure}
    \centering
    \includestandalone[mode=buildnew]{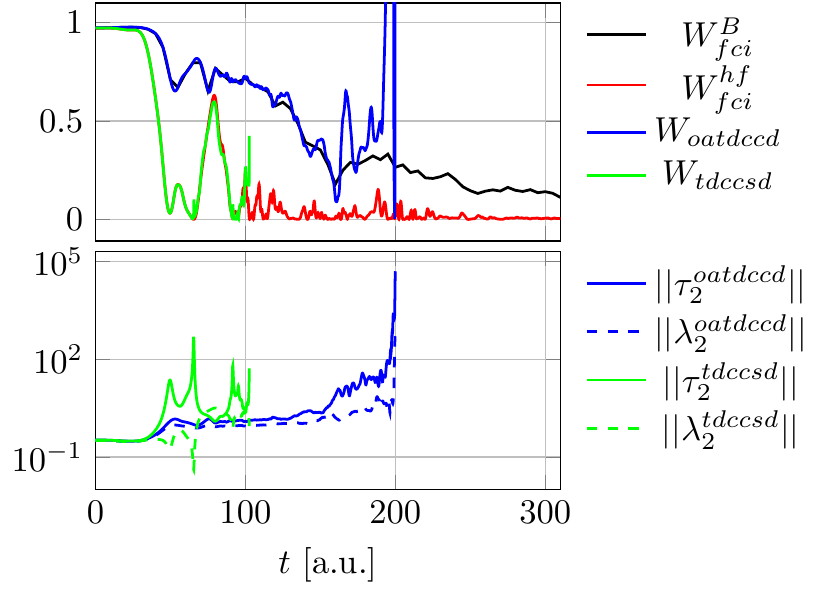}
    \caption{TDCCSD, OATDCCD and TDFCI simulations of \ch{LiH} with the cc-pVDZ basis exposed to a laser pulse of with $E_{\text{max}}=1$ a.u., $\omega = 0.06$ a.u. and $\phi = \pi/2$.
    }
    \label{reference weights lih cc-pVDZ breakdown}
\end{figure}

We thus see that OATDCCD theory offers improved numerical stability compared with TDCCSD theory.
Also OATDCCD theory should become unstable if the reference weight becomes greater than $1$. Attempting to trigger such a situation, we have performed an OATDCCD simulation of \ch{LiH} with the cc-pVDZ basis set and laser pulse parameters $E_{\text{max}} = 1\,\text{a.u.}$, $\omega = 0.06\,\text{a.u.}$,  $\phi = \pi/2$, and $t_d = 6\pi/\omega\,\text{a.u.}$. The $10$-fold increase in electric-field strength corresponds to a $100$-fold increase in laser intensity, which causes both the TDCCSD and OATDCCD methods to fail, see Fig.~\ref{reference weights lih cc-pVDZ breakdown}. 
The OATDCCD reference weight becomes greater than $1$ around $t=190\,\text{a.u.}$. As discussed above, this can be viewed as
a tell-tale sign of a poor CC approximation of the TDFCI wave function and, indeed, we observe that the accuracy of the OATDCCD dipole moment deteriorates from this point. 

\section*{Concluding remarks}
Numerical experiments demonstrate that the OATDCCD method provides a more stable approximation than TDCCSD theory for the description of laser-driven many-electron dynamics. Although also the OATDCCD method may be destabilized, it requires significantly higher field strengths to do so.
With less extreme laser pulses, dipole moments computed at the OATDCCD level of theory agree well with TDFCI results throughout the dynamics, with errors on the same order of magnitude as those of TDCCSD theory.
\added{This calls for further investigations of OATDCC theory, both for quantum dynamics and for response theory of molecular properties.}

We ascribe the enhanced numerical stability of OATDCCD theory to the use of optimal time-dependent biorthonormal reference determinants, which approximate the Brueckner determinant and its conjugate.  From this observation, we propose a simple diagnostic: the reference weight. Numerical difficulties must be expected when the reference weight approaches $0$ (for TDCCSD theory) or increases beyond $1$ (for OATDCCD theory).

\begin{acknowledgments}
This work was supported by the Research Council of Norway (RCN) through its Centres of Excellence scheme, project number 262695,
by the RCN Research Grant No. 240698,
and by the European Research Council under the European Union Seventh
Framework Program through the Starting Grant BIVAQUM, ERC-STG-2014 grant
agreement No 639508.
Support from the Norwegian Supercomputing Program (NOTUR) through a grant of computer time (Grant No.\ NN4654K) is
gratefully acknowledged.
\end{acknowledgments}

\bibliography{refs} 
\bibliographystyle{aipnum4-1}

\end{document}


\title{Numerical stability of time-dependent coupled-cluster methods for many-electron dynamics
in intense laser pulses}

\author{H{\aa}kon Emil Kristiansen}
\email{h.e.kristiansen@kjemi.uio.no}
\affiliation{Hylleraas Centre for Quantum Molecular Sciences,
Department of Chemistry, University of Oslo,
P.O. Box 1033 Blindern, N-0315 Oslo, Norway}

\author{{\O}yvind Sigmundson Sch{\o}yen}
\email{o.s.schoyen@fys.uio.no}
\affiliation{Department of Physics, University of Oslo,
N-0316 Oslo, Norway}

\author{Simen Kvaal}
\email{simen.kvaal@kjemi.uio.no}
\affiliation{Hylleraas Centre for Quantum Molecular Sciences,
Department of Chemistry, University of Oslo,
P.O. Box 1033 Blindern, N-0315 Oslo, Norway}

\author{Thomas Bondo Pedersen}
\email{t.b.pedersen@kjemi.uio.no}
\affiliation{Hylleraas Centre for Quantum Molecular Sciences,
Department of Chemistry, University of Oslo,
P.O. Box 1033 Blindern, N-0315 Oslo, Norway}

\date{\today}

\begin{abstract}
    This document gives additional results to the main article.
    Included are demonstrations of non-collapsing OATDCCD and TDCCSD simulations
    of \ch{Be} in the cc-pVDZ basis, discussion of controllable error in the
    two-particle case for TDFCI and OATDCCD, \ch{Be} in the cc-pVTZ basis, and
    systematic breaking of the TDCCSD and OATDCCD methods for \ch{LiH} in the
    aug-cc-pVDZ basis by increasing the field strength.
\end{abstract}

\maketitle

\section{\ch{Be} using TDCCSD}
\begin{figure}
        \centering
        \includestandalone[mode=buildnew]{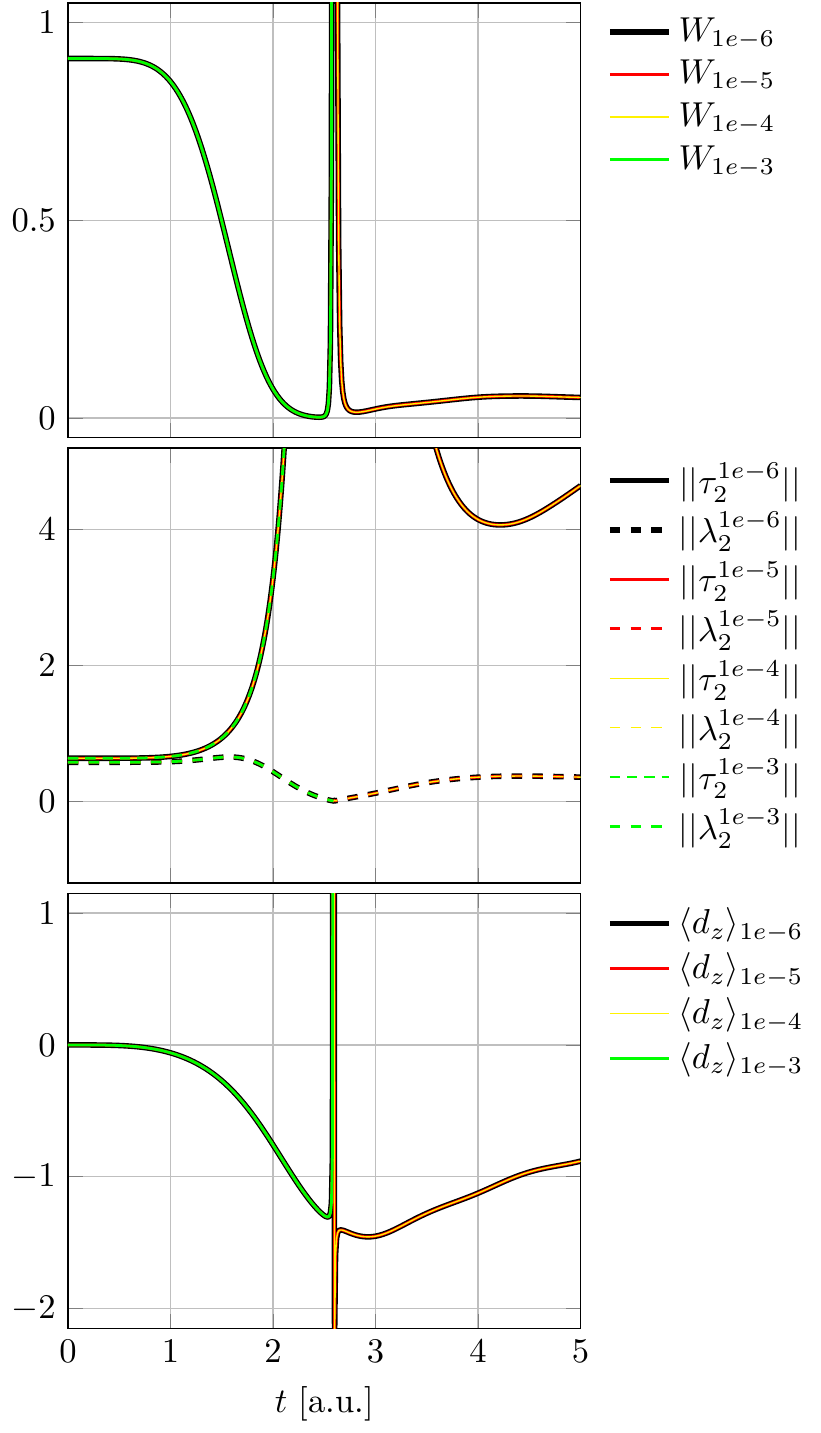}
        \caption{Reference weights, amplitude norms and induced dipole moment for
        \ch{Be} with the cc-pVDZ basis using TDCCSD with time steps $\Delta t \in
        \{10^{-3},\cdots, 10^{-6}\}\,\text{a.u.}$.}
        \label{stable tdccsd beryllium}
    \end{figure}

    Figure \ref{stable tdccsd beryllium} shows the TDCCSD reference weight, amplitude norms,
    and induced dipole moment for \ch{Be} with the cc-pVDZ basis.
    The maximum field strength is set to $E_{\text{max}}=1\,\text{a.u.}$, the carrier frequency of the laser 
    pulse is set to $\omega=0.2068175\,\text{a.u.}$ with phase $\phi=\pi/2$.
    The order parameter of the Gauss-Legendre integrator 
    is set to $s=3$ and the convergence threshold is $\epsilon=10^{-5}$. The time step is varied from $\Delta t = 10^{-3}\,\text{a.u.}$ to 
    $\Delta t = 10^{-6}\,\text{a.u.}$. Note that for $\Delta t = 10^{-3}\,\text{a.u.}$ the simulation breaks down at roughly $t=2.5\,\text{a.u.}$.
    With even smaller time steps, the simulations manage to complete, but not without the very noticeable
    spike in all measured quantities.

\section{Dipole error for \ch{He} in the cc-pVDZ basis}
    For two particles OATDCCD is formally equivalent to TDFCI. In the main paper 
    we report a maximum discrepancy between the induced dipole moment for a \ch{He}
    simulation on the order of $10^{-5}\,\text{a.u.}$. We repeat the simulation here, but with tighter 
    convergence thresholds and smaller time steps in the Gauss-Legendre integrator. 

\begin{figure}
        \centering
        \includestandalone[mode=buildnew]{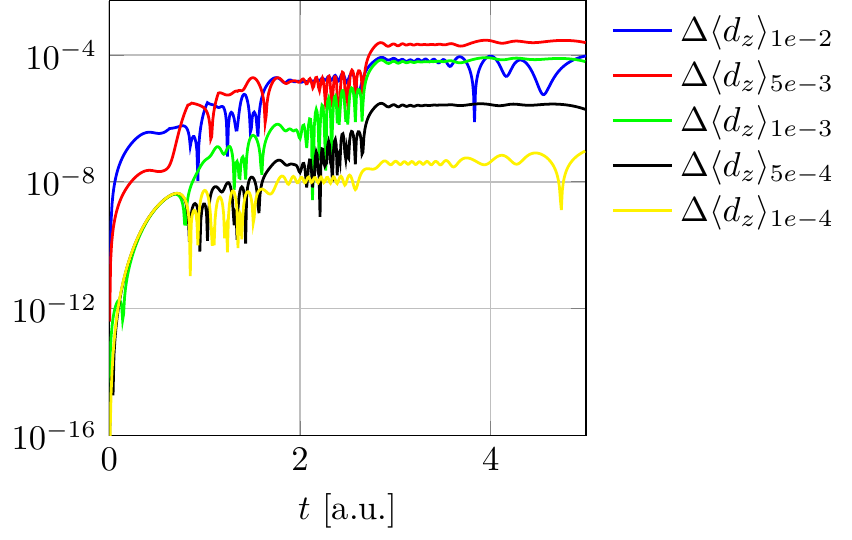}
        \caption{The absolute difference between the dipole moment of \ch{He} with
        cc-pVDZ basis from TDFCI and OATDCCD with $s = 3$, $\epsilon = 10^{-5}$
        for the Gauss-Legendre integrator, and ground-state convergence tolerance of OATDCCD
        of $10^{-5}$.}
        \label{he dipole error 1e-5}
    \end{figure}

    We keep $s = 3$ and vary $\epsilon$ and the ground-state convergence tolerance (in the OACCD ground-state solver) 
    over $\{10^{-5}, \cdots, 10^{-10} \}$ with $\Delta t \in \{10^{-2}, \cdots, 10^{-4} \}\,\text{a.u.}$. We note that for the FCI groundstate 
    we use a brute-force approach where the entire Hamiltonian is diagonalized, so there is no convergence parameter to be 
    specified. We compare the dipole moment from TDFCI and OATDCCD with the same parameters in both runs.
    The results are shown in Figs. \ref{he dipole error 1e-5} and \ref{he
    dipole error 1e-10}. The maximum error with $\Delta t = 10^{-4}\,\text{a.u.}$, ground-state convergence tolerance $10^{-10}$, and $\epsilon=10^{-10}$ 
    is $3.90499 \cdot 10^{-10}\,\text{a.u.}$. We thus see that the agreement between OATDCCD and TDFCI becomes smaller for tighter integration
    and ground-state convergence parameters. 

    \begin{figure}
        \centering
        \includestandalone[mode=buildnew]{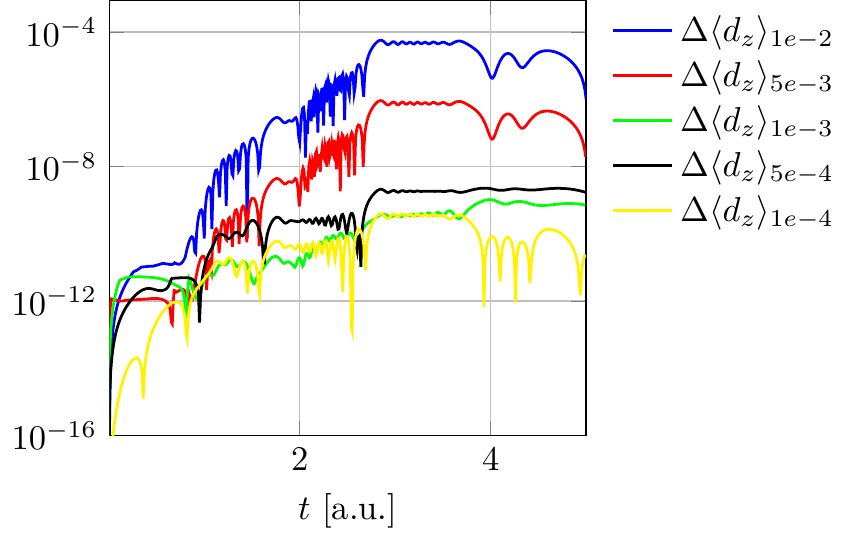}
        \caption{The absolute difference between the dipole moment of \ch{He} with
        cc-pVDZ basis from TDFCI and OATDCCD with $s = 3$, $\epsilon = 10^{-10}$
        for the Gauss-Legendre integrator, and ground-state convergence tolerance of OATDCCD
        of $10^{-10}$.}
        \label{he dipole error 1e-10}
    \end{figure}

\section{\ch{Be} with the cc-pVTZ basis}
\begin{figure}
        \centering
        \includestandalone[mode=buildnew]{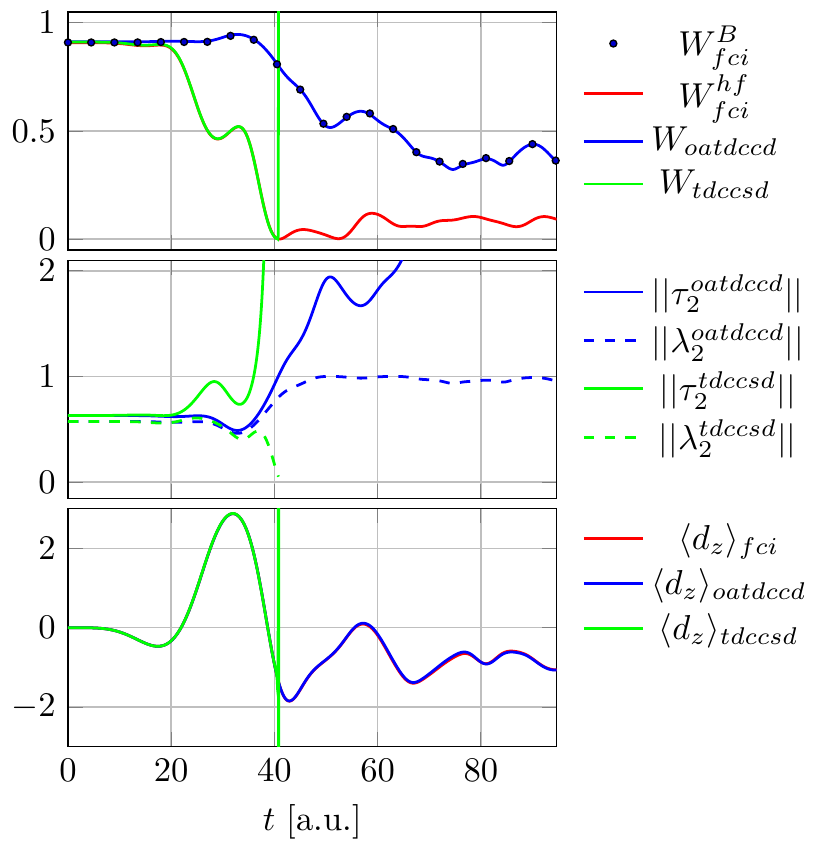}
        \caption{Reference weights, amplitude norms, and induced dipole moment for
        \ch{Be} with the cc-pVTZ basis.}
        \label{reference weights beryllium cc-pvtz}
    \end{figure}

Figure \ref{reference weights beryllium cc-pvtz} shows the reference weights, amplitude norms, and the induced dipole moment for 
\ch{Be} with the cc-pVTZ basis, computed with TDFCI, TDCCSD and OATDCCD. The
maximum field strength is $E_{\text{max}} = 0.1\,\text{a.u.}$, the carrier frequency
$\omega=0.1990\,\text{a.u.}$, and the phase $\phi = 0$.  

\section{\ch{LiH} with aug-cc-pVDZ basis}
\begin{figure*}
        \centering
        \includestandalone[mode=buildnew]{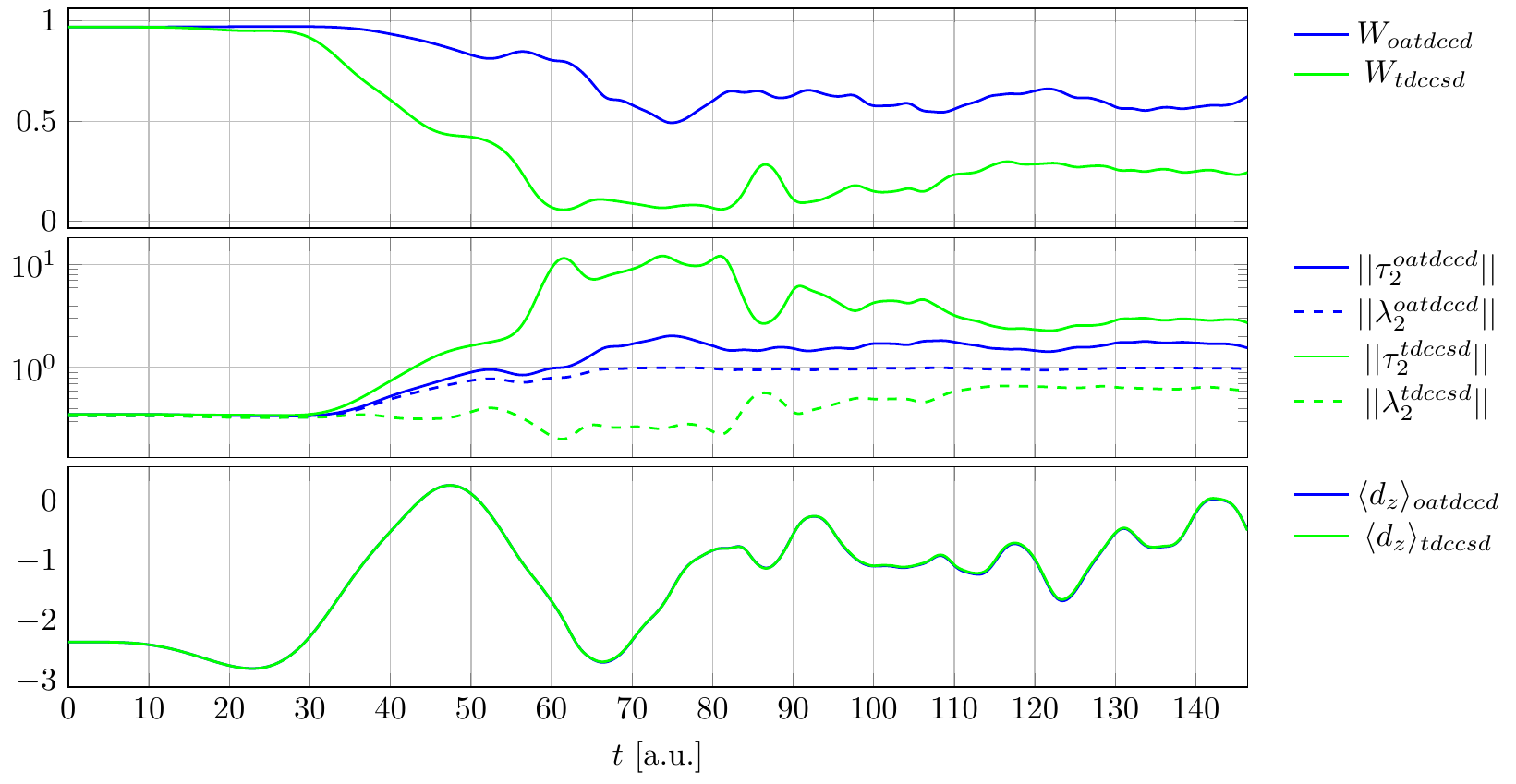}
        \caption{OATDCCD and TDCCSD simulations of \ch{LiH} with the aug-cc-pVDZ
        basis using a field strength of $E_{\text{max}} =
        0.1\left(\sqrt{2}\right)\,\text{a.u.}$.}
        \label{breakdown lih 1}
    \end{figure*}

\begin{figure*}
        \centering
        \includestandalone[mode=buildnew]{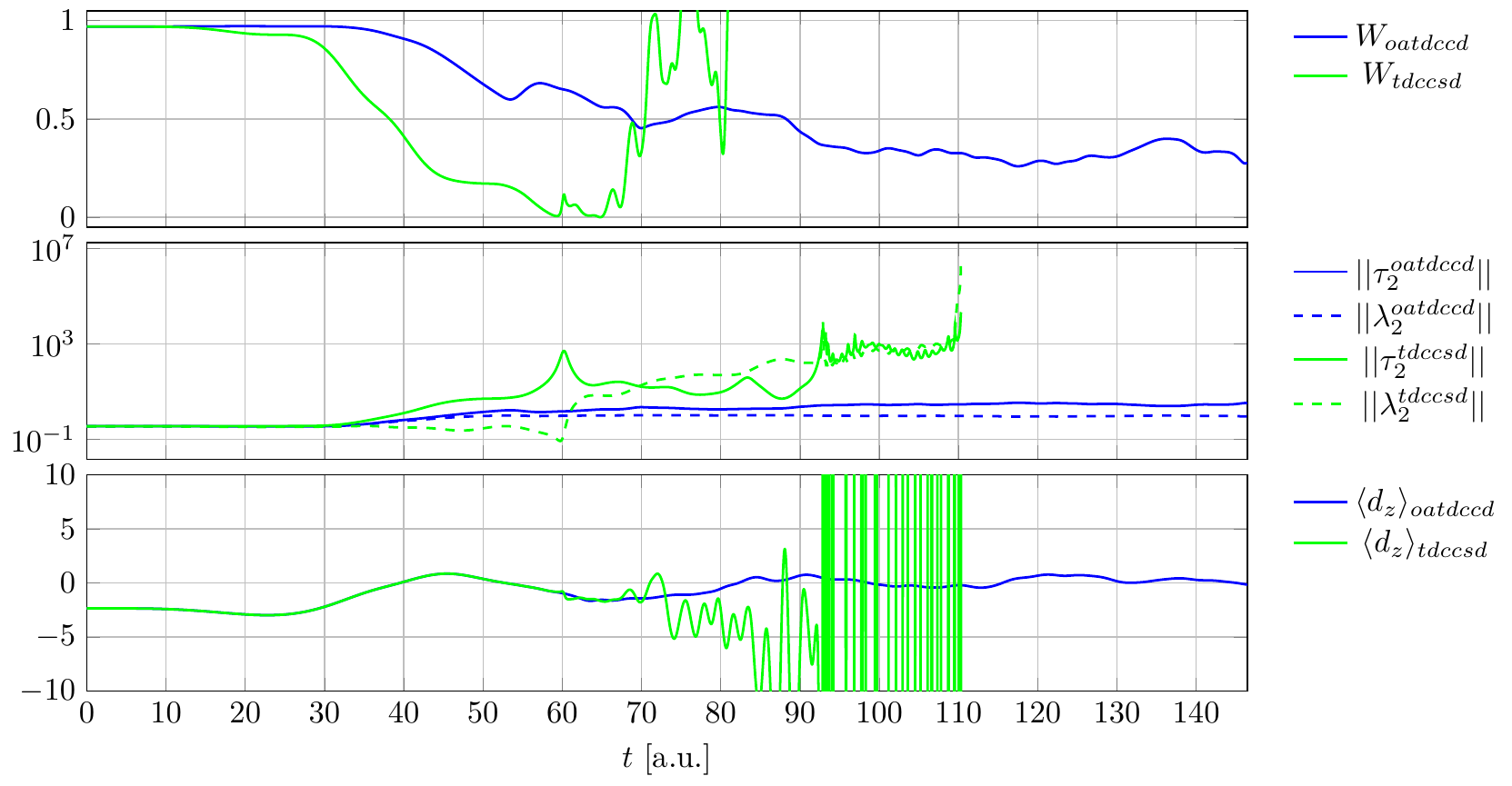}
        \caption{OATDCCD and TDCCSD simulations of \ch{LiH} with the aug-cc-pVDZ
        basis using a field strength of $E_{\text{max}} =
        0.1\left(\sqrt{2}\right)^2\,\text{a.u.}$.}
        \label{breakdown lih 2}
    \end{figure*}

\begin{figure*}
        \centering
        \includestandalone[mode=buildnew]{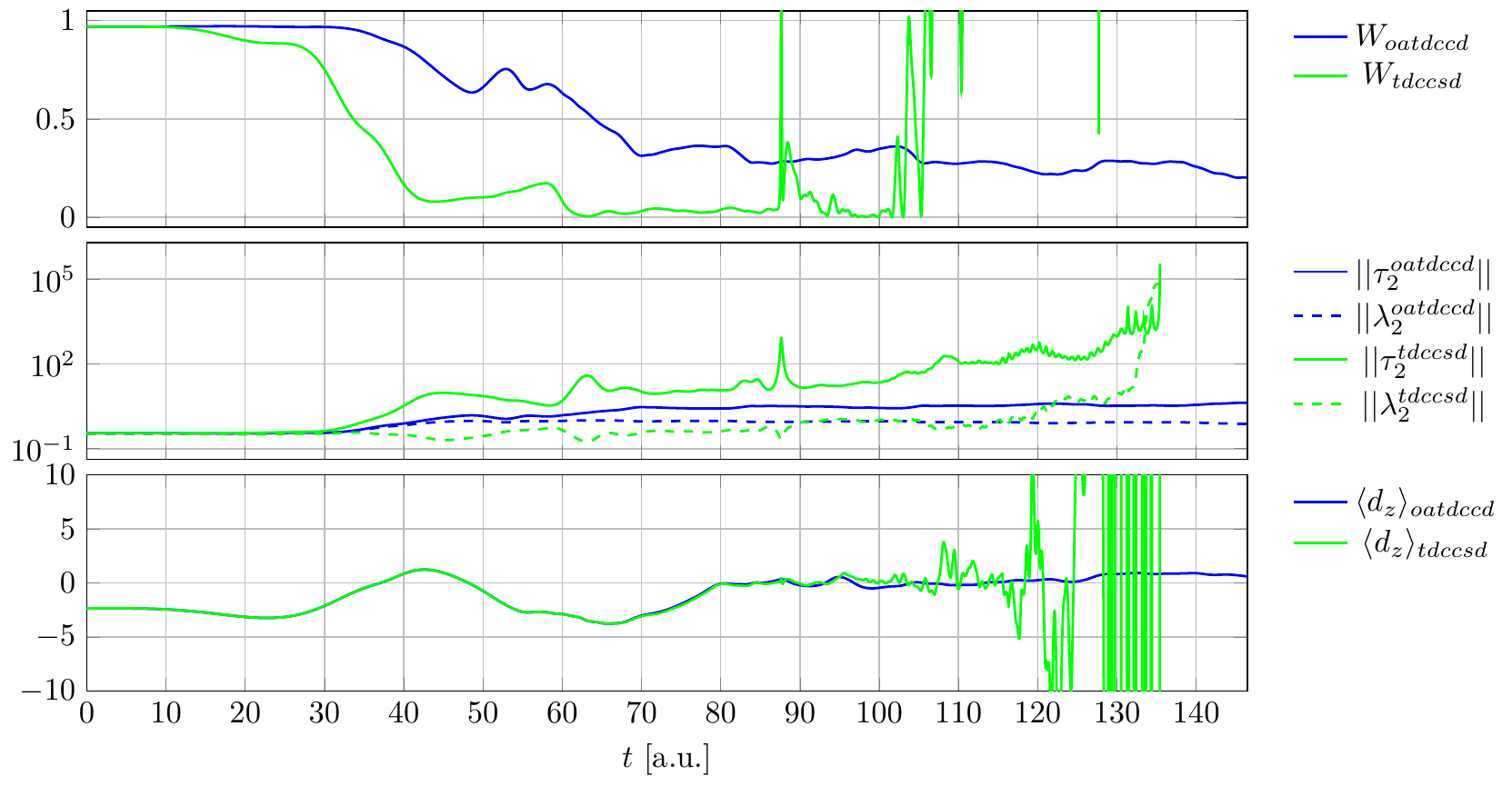}
        \caption{OATDCCD and TDCCSD simulations of \ch{LiH} with the aug-cc-pVDZ
        basis using a field strength of $E_{\text{max}} =
        0.1\left(\sqrt{2}\right)^3\,\text{a.u.}$.}
        \label{breakdown lih 3}
    \end{figure*}

\begin{figure*}
        \centering
        \includestandalone[mode=buildnew]{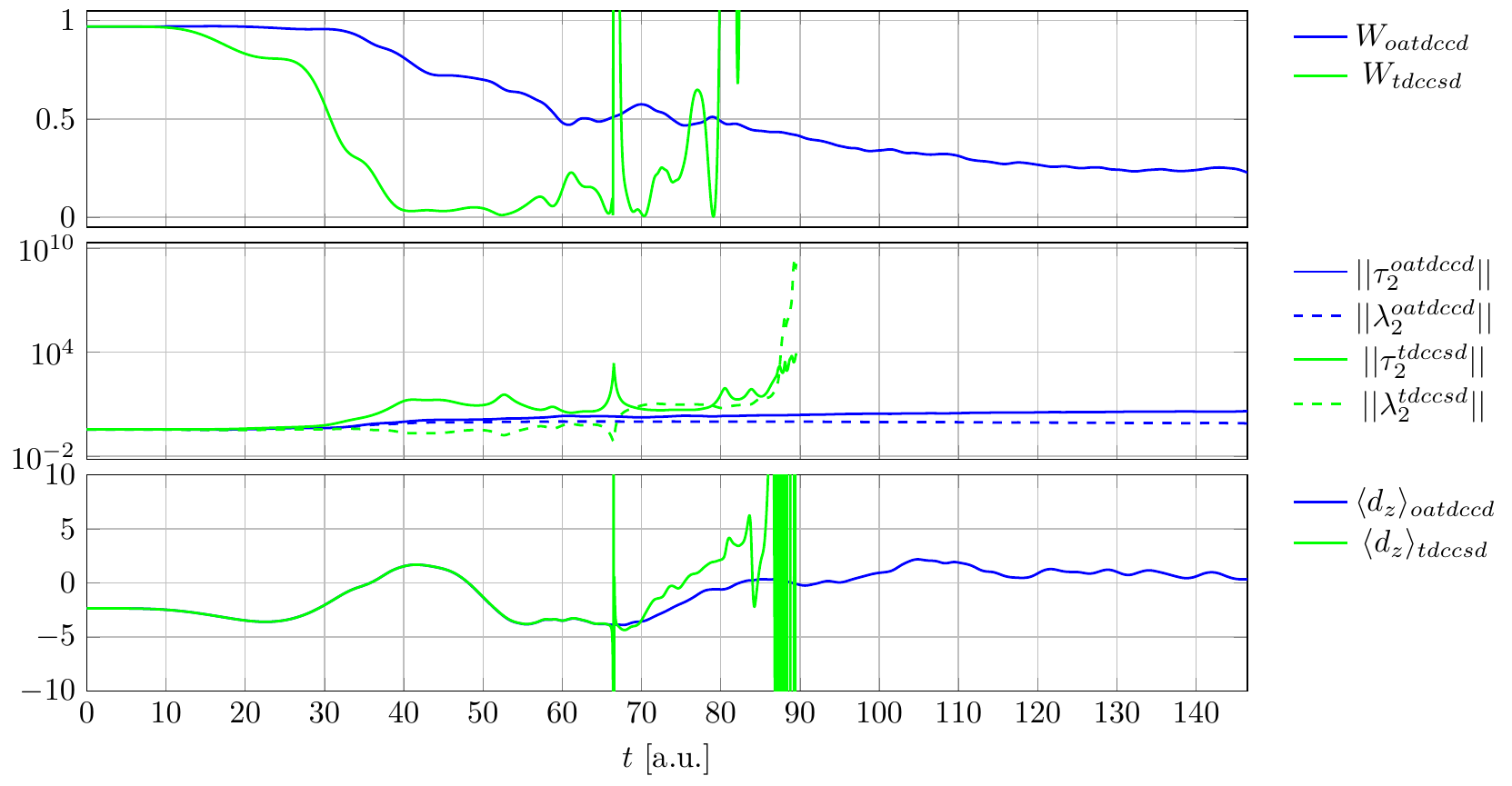}
        \caption{OATDCCD and TDCCSD simulations of \ch{LiH} with the aug-cc-pVDZ
        basis using a field strength of $E_{\text{max}} =
        0.1\left(\sqrt{2}\right)^4\,\text{a.u.}$.}
        \label{breakdown lih 4}
    \end{figure*}

\begin{figure*}
        \centering
        \includestandalone[mode=buildnew]{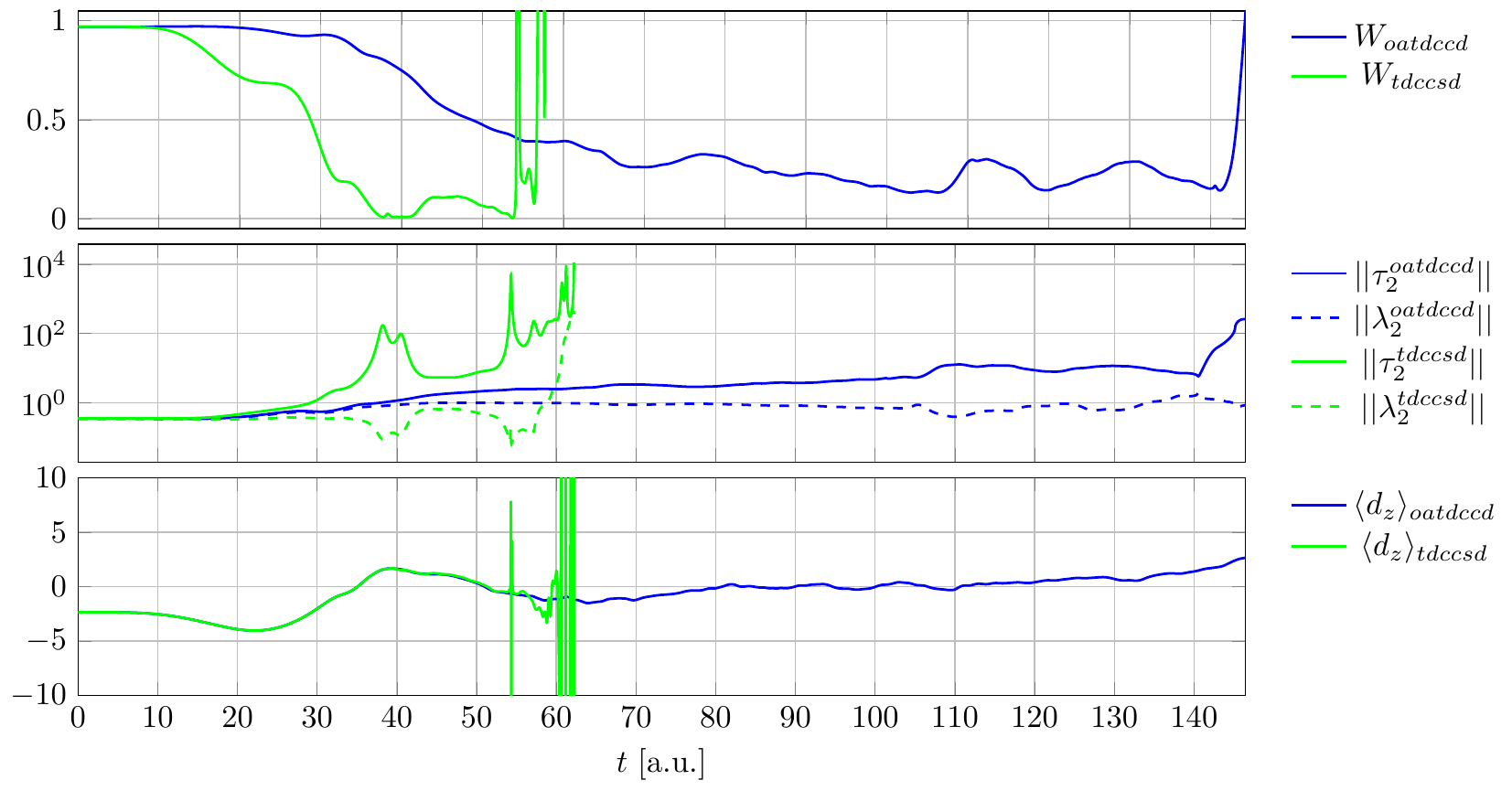}
        \caption{OATDCCD and TDCCSD simulations of \ch{LiH} with the aug-cc-pVDZ
        basis using a field strength of $E_{\text{max}} =
        0.1\left(\sqrt{2}\right)^5\,\text{a.u.}$.}
        \label{breakdown lih 5}
    \end{figure*}

\begin{figure*}
        \centering
        \includestandalone[mode=buildnew]{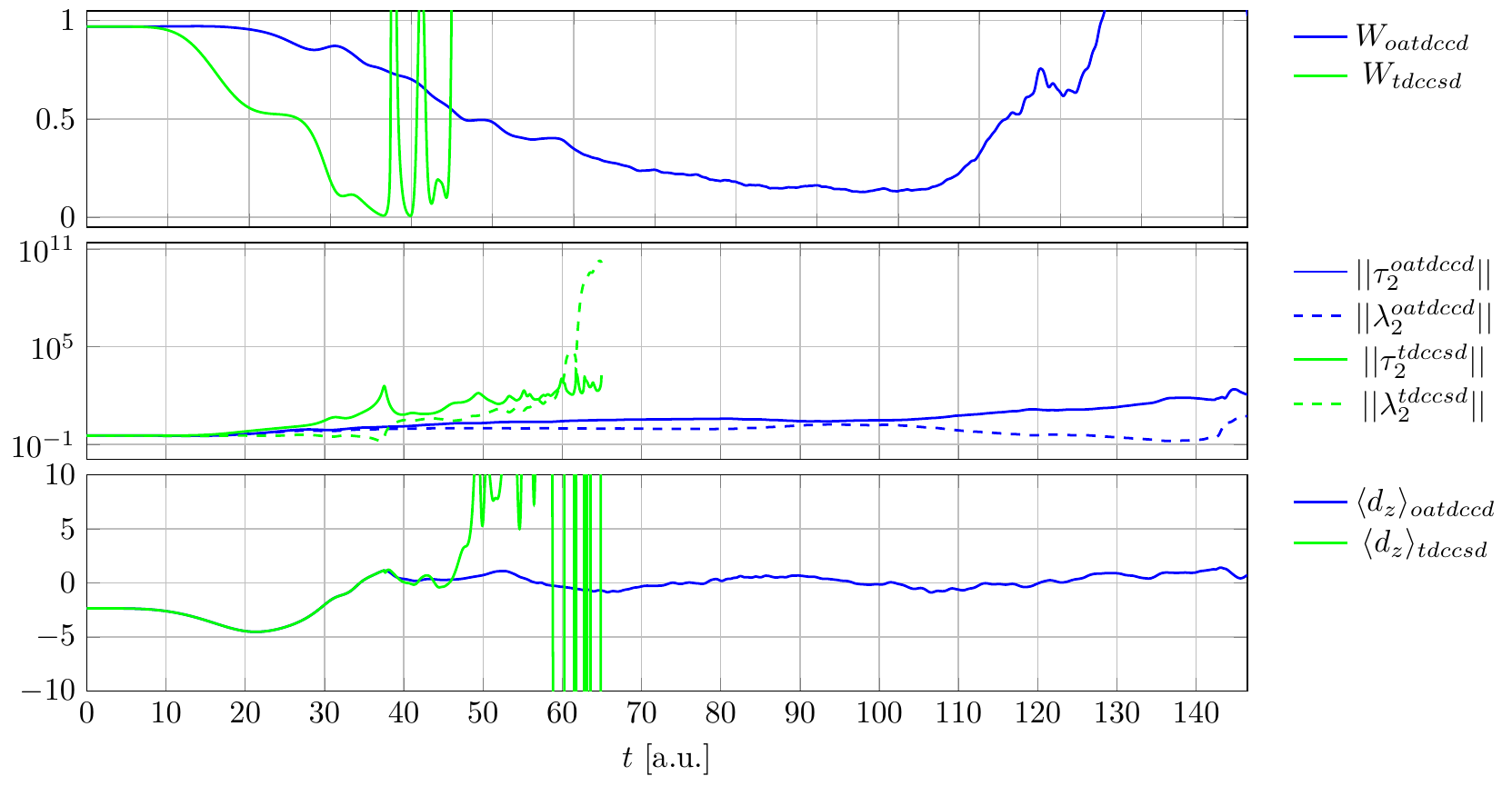}
        \caption{OATDCCD and TDCCSD simulations of \ch{LiH} with the aug-cc-pVDZ
        basis using a field strength of $E_{\text{max}} =
        0.1\left(\sqrt{2}\right)^6\,\text{a.u.}$.}
        \label{breakdown lih 6}
    \end{figure*}

\begin{figure*}
        \centering
        \includestandalone[mode=buildnew]{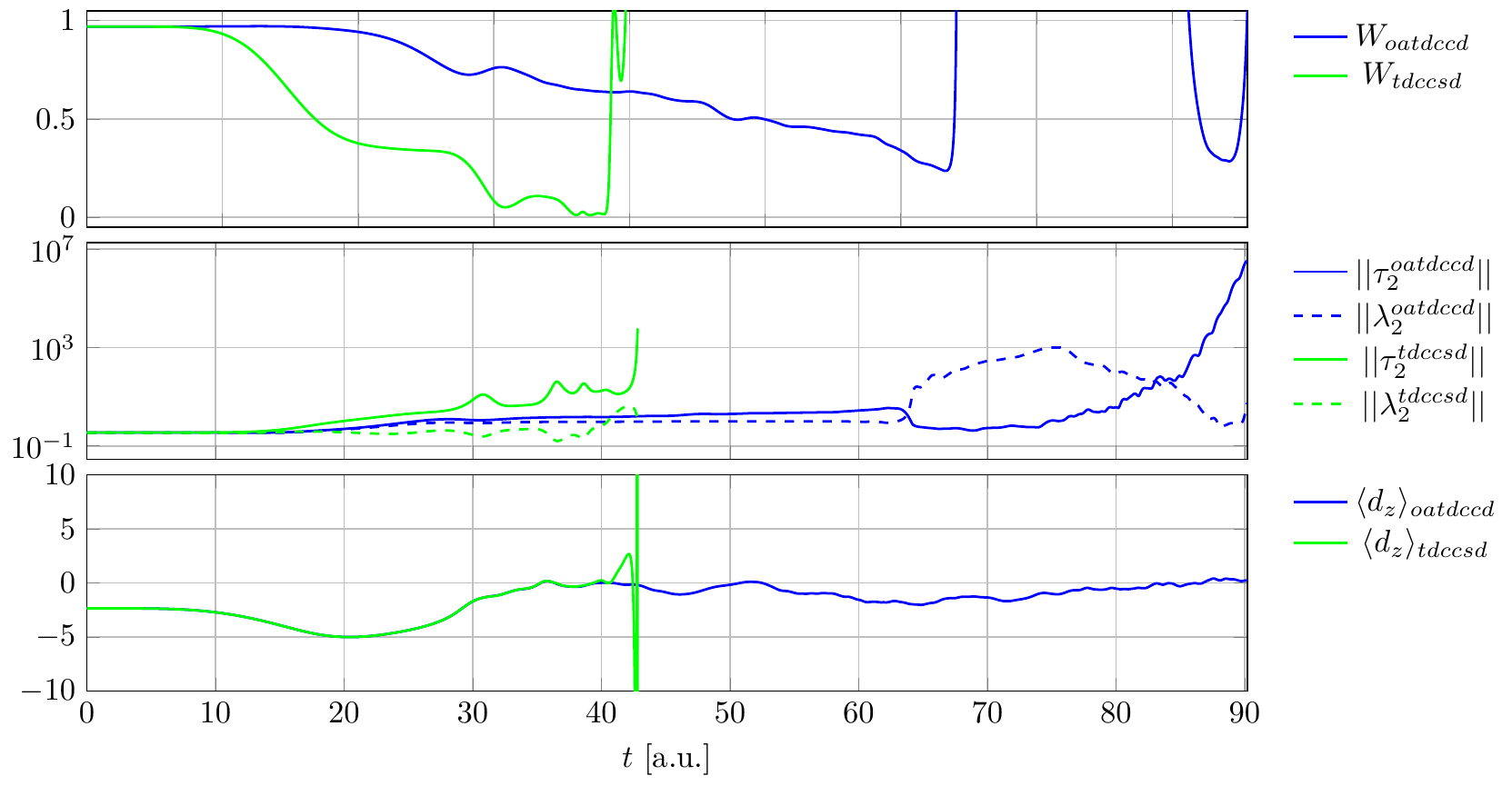}
        \caption{OATDCCD and TDCCSD simulations of \ch{LiH} with the aug-cc-pVDZ
        basis using a field strength of $E_{\text{max}} =
        0.1\left(\sqrt{2}\right)^7\,\text{a.u.}$.}
        \label{breakdown lih 7}
    \end{figure*}

In order to investigate further how strong fields can be handled by TDCCSD and
OATDCCD, Figs. \ref{breakdown lih 1}-\ref{breakdown lih 7} 
show the reference weights and induced dipole moment for \ch{LiH} with the
aug-cc-pVDZ basis. The carrier frequency of the laser pulse is set to
$\omega=0.1287\,\text{a.u.}$ with phase $\phi=0$.
The field strength is increased in powers of $\sqrt{2}$, corresponding to
successive doubling of the intensity of the laser.
The TDCCSD method breaks down at $E_{\text{max}} = 0.2\,\text{a.u.}$.
The OATDCCD method breaks down at $E_{\text{max}} = 0.1(\sqrt{2})^5\,\text{a.u.}$ where
the reference weight becomes greater than one at the end of the simulation. 

\bibliography{refs} 
\bibliographystyle{aipnum4-1}